\documentclass[RNAAS, modern]{aastex63}

\begin{document}

\title{VTSCat: The VERITAS Catalog of Gamma-Ray Observations}

\author{A.~Acharyya}\affiliation{Department of Physics and Astronomy, University of Alabama, Tuscaloosa, AL 35487, USA}
\author{C.~B.~Adams}\affiliation{Physics Department, Columbia University, New York, NY 10027, USA}
\author{A.~Archer}\affiliation{Department of Physics and Astronomy, DePauw University, Greencastle, IN 46135-0037, USA}
\author{P.~Bangale}\affiliation{Department of Physics and Astronomy and the Bartol Research Institute, University of Delaware, Newark, DE 19716, USA}
\author{J.~T.~Bartkoske}\affiliation{Department of Physics and Astronomy, University of Utah, Salt Lake City, UT 84112, USA}
\author{P.~Batista}\affiliation{DESY, Platanenallee 6, 15738 Zeuthen, Germany}
\author{W.~Benbow}\affiliation{Center for Astrophysics, Harvard \& Smithsonian, Cambridge, MA 02138, USA}
\author{A.~Brill}\affiliation{N.A.S.A./Goddard Space-Flight Center, Code 661, Greenbelt, MD 20771, USA}
\author{R.~Brose}\affiliation{Institute of Physics and Astronomy, University of Potsdam, 14476 Potsdam-Golm, Germany and DESY, Platanenallee 6, 15738 Zeuthen, Germany}
\author{J.~H.~Buckley}\affiliation{Department of Physics, Washington University, St. Louis, MO 63130, USA}
\author{M.~Capasso}\affiliation{Department of Physics and Astronomy, Barnard College, Columbia University, NY 10027, USA}
\author{J.~L.~Christiansen}\affiliation{Physics Department, California Polytechnic State University, San Luis Obispo, CA 94307, USA}
\author{A.~J.~Chromey}\affiliation{Center for Astrophysics, Harvard \& Smithsonian, Cambridge, MA 02138, USA}
\author{M.~K.~Daniel}\affiliation{Center for Astrophysics, Harvard \& Smithsonian, Cambridge, MA 02138, USA}
\author{M.~Errando}\affiliation{Department of Physics, Washington University, St. Louis, MO 63130, USA}
\author{A.~Falcone}\affiliation{Department of Astronomy and Astrophysics, 525 Davey Lab, Pennsylvania State University, University Park, PA 16802, USA}
\author{K.~A~Farrell}\affiliation{School of Physics, University College Dublin, Belfield, Dublin 4, Ireland}
\author{Q.~Feng}\affiliation{Center for Astrophysics, Harvard \& Smithsonian, Cambridge, MA 02138, USA}
\author{J.~P.~Finley}\affiliation{Department of Physics and Astronomy, Purdue University, West Lafayette, IN 47907, USA}
\author{G.~M~Foote}\affiliation{Department of Physics and Astronomy and the Bartol Research Institute, University of Delaware, Newark, DE 19716, USA}
\author{L.~Fortson}\affiliation{School of Physics and Astronomy, University of Minnesota, Minneapolis, MN 55455, USA}
\author{A.~Furniss}\affiliation{Department of Physics, California State University - East Bay, Hayward, CA 94542, USA}
\author{G.~Gallagher}\affiliation{Department of Physics and Astronomy, Ball State University, Muncie, IN 47306, USA}
\author{A.~Gent}\affiliation{School of Physics and Center for Relativistic Astrophysics, Georgia Institute of Technology, 837 State Street NW, Atlanta, GA 30332-0430}
\author{C.~Giuri}\affiliation{DESY, Platanenallee 6, 15738 Zeuthen, Germany}
\author{O.~Gueta}\affiliation{DESY, Platanenallee 6, 15738 Zeuthen, Germany}
\author{W.~F~Hanlon}\affiliation{Center for Astrophysics, Harvard \& Smithsonian, Cambridge, MA 02138, USA}
\author{D.~Hanna}\affiliation{Physics Department, McGill University, Montreal, QC H3A 2T8, Canada}
\author{T.~Hassan}\affiliation{DESY, Platanenallee 6, 15738 Zeuthen, Germany}
\author{O.~Hervet}\affiliation{Santa Cruz Institute for Particle Physics and Department of Physics, University of California, Santa Cruz, CA 95064, USA}
\author{J.~Hoang}\affiliation{Santa Cruz Institute for Particle Physics and Department of Physics, University of California, Santa Cruz, CA 95064, USA}
\author{J.~Holder}\affiliation{Department of Physics and Astronomy and the Bartol Research Institute, University of Delaware, Newark, DE 19716, USA}
\author{G.~Hughes}\affiliation{Center for Astrophysics, Harvard \& Smithsonian, Cambridge, MA 02138, USA}
\author{T.~B.~Humensky}\affiliation{Department of Physics, University of Maryland, College Park, MD, USA and }
\author{W.~Jin}\affiliation{Department of Physics and Astronomy, University of Alabama, Tuscaloosa, AL 35487, USA}
\author{P.~Kaaret}\affiliation{Department of Physics and Astronomy, University of Iowa, Van Allen Hall, Iowa City, IA 52242, USA}
\author{M.~Kertzman}\affiliation{Department of Physics and Astronomy, DePauw University, Greencastle, IN 46135-0037, USA}
\author{D.~Kieda}\affiliation{Department of Physics and Astronomy, University of Utah, Salt Lake City, UT 84112, USA}
\author{T.~K.~Kleiner}\affiliation{DESY, Platanenallee 6, 15738 Zeuthen, Germany}
\author{N.~Korzoun}\affiliation{Department of Physics and Astronomy and the Bartol Research Institute, University of Delaware, Newark, DE 19716, USA}
\author{F.~Krennrich}\affiliation{Department of Physics and Astronomy, Iowa State University, Ames, IA 50011, USA}
\author{S.~Kumar}\affiliation{Department of Physics, University of Maryland, College Park, MD, USA }
\author{M.~J.~Lang}\affiliation{School of Natural Sciences, University of Galway, University Road, Galway, H91 TK33, Ireland}
\author{M.~Lundy}\affiliation{Physics Department, McGill University, Montreal, QC H3A 2T8, Canada}
\author{G.~Maier}\affiliation{DESY, Platanenallee 6, 15738 Zeuthen, Germany}
\author{C.~E~McGrath}\affiliation{School of Physics, University College Dublin, Belfield, Dublin 4, Ireland}
\author{M.~J~Millard}\affiliation{Department of Physics and Astronomy, University of Iowa, Van Allen Hall, Iowa City, IA 52242, USA}
\author{C.~L.~Mooney}\affiliation{Department of Physics and Astronomy and the Bartol Research Institute, University of Delaware, Newark, DE 19716, USA}
\author{P.~Moriarty}\affiliation{School of Natural Sciences, University of Galway, University Road, Galway, H91 TK33, Ireland}
\author{R.~Mukherjee}\affiliation{Department of Physics and Astronomy, Barnard College, Columbia University, NY 10027, USA}
\author{D.~Nieto}\affiliation{Institute of Particle and Cosmos Physics, Universidad Complutense de Madrid, 28040 Madrid, Spain}
\author{M.~Nievas-Rosillo}\affiliation{DESY, Platanenallee 6, 15738 Zeuthen, Germany}
\author{S.~O'Brien}\affiliation{Physics Department, McGill University, Montreal, QC H3A 2T8, Canada}
\author{R.~A.~Ong}\affiliation{Department of Physics and Astronomy, University of California, Los Angeles, CA 90095, USA}
\author{A.~N.~Otte}\affiliation{School of Physics and Center for Relativistic Astrophysics, Georgia Institute of Technology, 837 State Street NW, Atlanta, GA 30332-0430}
\author{D.~Pandel}\affiliation{Department of Physics, Grand Valley State University, Allendale, MI 49401, USA}
\author{N.~Park}\affiliation{Department of Physics, Engineering Physics and Astronomy, Queen`s University, Kingston, ON K7L 3N6, Canada}
\author{S.~R.~Patel}\affiliation{DESY, Platanenallee 6, 15738 Zeuthen, Germany}
\author{S.~Patel}\affiliation{Department of Physics and Astronomy, University of Iowa, Van Allen Hall, Iowa City, IA 52242, USA}
\author{K.~Pfrang}\affiliation{DESY, Platanenallee 6, 15738 Zeuthen, Germany}
\author{A.~Pichel}\affiliation{Instituto de Astronomía y Física del Espacio (IAFE, CONICET-UBA), CC 67 - Suc. 28, (C1428ZAA) Ciudad Autónoma de Buenos Aires, Argentina}
\author{M.~Pohl}\affiliation{Institute of Physics and Astronomy, University of Potsdam, 14476 Potsdam-Golm, Germany and DESY, Platanenallee 6, 15738 Zeuthen, Germany}
\author{R.~R.~Prado}\affiliation{DESY, Platanenallee 6, 15738 Zeuthen, Germany}
\author{E.~Pueschel}\affiliation{DESY, Platanenallee 6, 15738 Zeuthen, Germany}
\author{J.~Quinn}\affiliation{School of Physics, University College Dublin, Belfield, Dublin 4, Ireland}
\author{K.~Ragan}\affiliation{Physics Department, McGill University, Montreal, QC H3A 2T8, Canada}
\author{P.~T.~Reynolds}\affiliation{Department of Physical Sciences, Munster Technological University, Bishopstown, Cork, T12 P928, Ireland}
\author{D.~Ribeiro}\affiliation{Physics Department, Columbia University, New York, NY 10027, USA}
\author{G.~T.~Richards}\affiliation{Department of Physics and Astronomy and the Bartol Research Institute, University of Delaware, Newark, DE 19716, USA}
\author{E.~Roache}\affiliation{Center for Astrophysics, Harvard \& Smithsonian, Cambridge, MA 02138, USA}
\author{A.~C.~Rovero}\affiliation{Instituto de Astronomía y Física del Espacio (IAFE, CONICET-UBA), CC 67 - Suc. 28, (C1428ZAA) Ciudad Autónoma de Buenos Aires, Argentina}
\author{C.~Rulten}\affiliation{School of Physics and Astronomy, University of Minnesota, Minneapolis, MN 55455, USA}
\author{J.~L.~Ryan}\affiliation{Department of Physics and Astronomy, University of California, Los Angeles, CA 90095, USA}
\author{I.~Sadeh}\affiliation{DESY, Platanenallee 6, 15738 Zeuthen, Germany}
\author{M.~Santander}\affiliation{Department of Physics and Astronomy, University of Alabama, Tuscaloosa, AL 35487, USA}
\author{S.~Schlenstedt}\affiliation{CTAO, Saupfercheckweg 1, 69117 Heidelberg, Germany}
\author{G.~H.~Sembroski}\affiliation{Department of Physics and Astronomy, Purdue University, West Lafayette, IN 47907, USA}
\author{R.~Shang}\affiliation{Department of Physics and Astronomy, University of California, Los Angeles, CA 90095, USA}
\author{M.~Splettstoesser}\affiliation{Santa Cruz Institute for Particle Physics and Department of Physics, University of California, Santa Cruz, CA 95064, USA}
\author{B.~Stevenson}\affiliation{Department of Physics and Astronomy, University of California, Los Angeles, CA 90095, USA}
\author{D.~Tak}\affiliation{DESY, Platanenallee 6, 15738 Zeuthen, Germany}
\author{V.~V.~Vassiliev}\affiliation{Department of Physics and Astronomy, University of California, Los Angeles, CA 90095, USA}
\author{S.~P.~Wakely}\affiliation{Enrico Fermi Institute, University of Chicago, Chicago, IL 60637, USA}
\author{A.~Weinstein}\affiliation{Department of Physics and Astronomy, Iowa State University, Ames, IA 50011, USA}
\author{D.~A.~Williams}\affiliation{Santa Cruz Institute for Particle Physics and Department of Physics, University of California, Santa Cruz, CA 95064, USA}
\author{T.~J~Williamson}\affiliation{Department of Physics and Astronomy and the Bartol Research Institute, University of Delaware, Newark, DE 19716, USA}

\author{L.~Angelini}\affiliation{NASA Goddard Space Flight Center, Code 662, Greenbelt, MD 20771, USA}
\author{A.~Basu-Zych}\affiliation{NASA Goddard Space Flight Center, Code 662, Greenbelt, MD 20771, USA}\affiliation{Center for Space Science and Technology, University of Maryland Baltimore County, 1000 Hilltop Circle, Baltimore, MD 21250, USA}
\author{E.~Sabol}\affiliation{NASA Goddard Space Flight Center, Code 662, Greenbelt, MD 20771, USA}\affiliation{Center for Space Science and Technology, University of Maryland Baltimore County, 1000 Hilltop Circle, Baltimore, MD 21250, USA}
\author{A.~Smale}\affiliation{ADNET Systems Inc., 6720B Rockledge Drive, Suite \#504, Bethesda, MD 20817}

\correspondingauthor{Gernot Maier}
\email{gernot.maier@desy.de}
\correspondingauthor{Sameer Patel}
\email{sameer-patel-1@uiowa.edu}
\correspondingauthor{Philip Kaaret}
\email{philip-kaaret@uiowa.edu}



%



\begin{abstract}
The ground-based gamma-ray observatory VERITAS (Very Energetic Radiation Imaging Telescope Array System\footnote{\url{https://veritas.sao.arizona.edu/}}, \citep{2006APh....25..391H}) is sensitive to photons of astrophysical origin with energies in the range between $\approx 85$ GeV to $\approx 30$ TeV.
The instrument  consists of four 12-m diameter imaging Cherenkov telescopes operating at the Fred Lawrence Whipple Observatory (FLWO) in southern Arizona.
VERITAS started four-telescope operations in 2007 and collects about 1100 hours of good-weather data per year.
The VERITAS collaboration has published over 100 journal articles since 2008 reporting on gamma-ray observations of a large
variety of objects: Galactic sources like supernova remnants, pulsar wind nebulae, and binary systems; 
extragalactic sources like star forming galaxies, dwarf-spheroidal galaxies, and highly-variable active galactic nuclei.
This note presents VTSCat: the catalog of high-level data products from all VERITAS publications.
\end{abstract}

\section{VTSCat}

VTSCat is inspired by various movements for open data, among them the initiative for data formats in gamma-ray astronomy\footnote{see \url{https://gamma-astro-data-formats.readthedocs.io/en/latest/}} \citep{2017AIPC.1792g0006D}. 
It is built on an early version of the gamma-ray catalog gamma-cat\footnote{see \url{https://gamma-cat.readthedocs.io/index.html}} and profited significantly from the input of the gamma-cat developers.
All VTSCat data can be accessed  file-by-file on GitHub (\url{https://github.com/VERITAS-Observatory/VERITAS-VTSCat}) or downloaded via Zenodo 
(\cite{benbow_w_2022_6163391}, downloadable as a compressed file).
The light curves and spectral results in VTSCat are also available through NASA's High Energy Astrophysics Science Archive Research Center (HEASARC; \url{https://heasarc.gsfc.nasa.gov/W3Browse/all/verimaster.html}), thus providing an interface more familiar to astronomers (see next section).

The VTSCat data collection contains: 
\begin{itemize}
\item  high-level data such as spectral flux points, light curves, spectral fits in human- and machine-readable \textit{yaml} and \textit{ecsv} file formats\footnote{\url{https://yaml.org/} and \url{https://github.com/astropy/astropy-APEs/blob/master/APE6.rst}},
\item table data such as upper limits from dark matter searches or results on the extragalactic background light in the \textit{ecsv} file format,
\item sky maps (wherever available) in the FITS file format.
\end{itemize}
The physical units are explicitly given in each file; the \textit{ecsv}-files follow standards which allow them to be read by common tools such as the astropy python library \citep{2022ApJ...935..167A}.

The data files are organized in VTSCat by year and publication, using the ADS bibcodes as reference identifiers. 
Objects are identified by a running integer (labeled \textit{source\_id} in data files) following the scheme developed by gamma-cat.
The description files for objects can be found in the \href{https://github.com/VERITAS-Observatory/VERITAS-VTSCat/tree/main/sources}{sources} subdirectory and include the most relevant names for a given object (common name in the field, VERITAS object identifier, primary identifier by SIMBAD), and the object coordinates. 

For illustration, the data entry for the VERITAS publication \cite{2018ApJ...861L..20A} is discussed in the following (see the directory \href{https://github.com/VERITAS-Observatory/VERITAS-VTSCat/tree/main/2018/2018ApJ...861L..20A}{2018/2018ApJ...861L..20A} on the repository to follow the discussion below).
A \href{https://github.com/VERITAS-Observatory/VERITAS-VTSCat/blob/main/2018/2018ApJ...861L..20A/README.md}{README} file provides an overview of all data products, figures, source names, and citations.
For this example, the data files are:

\begin{itemize}

\item Observation data: 
\href{https://github.com/VERITAS-Observatory/VERITAS-VTSCat/blob/main/2018/2018ApJ...861L..20A/VER-000168.yaml}{VER-000168.yaml}

Observational details and results like observation time, detection significance, and spectral models. 

\item Spectral flux points: \href{https://github.com/VERITAS-Observatory/VERITAS-VTSCat/blob/main/2018/2018ApJ...861L..20A/VER-000168-sed-1.ecsv}{VER-000168-sed-1.ecsv}

Spectral flux points including errors and upper limits as published in Fig 2 of \cite{2018ApJ...861L..20A}.

\item Light-curve data: \href{https://github.com/VERITAS-Observatory/VERITAS-VTSCat/blob/main/2018/2018ApJ...861L..20A/VER-000168-lc.ecsv}{VER-000168-lc.ecsv}

Light-curve data (integrated gamma-ray fluxes vs time) as published in Fig 3 of \cite{2018ApJ...861L..20A}.

\item FITS data: \href{https://github.com/VERITAS-Observatory/VERITAS-VTSCat/blob/main/2018/2018ApJ...861L..20A/VER-000168-signif-skymap.fits}{VER-000168-signif-skymap.fits}

Statistical significance sky map as published in Fig 1 of \cite{2018ApJ...861L..20A} \href{https://github.com/VERITAS-Observatory/VERITAS-VTSCat/blob/main/2018/2018ApJ...861L..20A/VER-000168-signif-skymap.fits}.
\end{itemize}

The yaml datatypes and spectral models used are explained in the documents  \href{https://github.com/VERITAS-Observatory/VERITAS-VTSCat/blob/main/Formats_and_Models.md}{Formats\_and\_Models.md}
and \href{https://github.com/VERITAS-Observatory/VERITAS-VTSCat/blob/main/SpectralModels.md}{SpectralModels.md}.
The data types and units used in \textit{ecsv} files are described at the top of each data file;
for FITS files this information is provided in the FITS header.
Preview PNG images are included for all lightcurves and spectral energy distribution plots.

Translation of the data in the publications to a uniform catalog was a complex process, requiring significant manual effort and occasional addition of information.
For example, we note the addition of the source name VER J1746-286 that refers to the diffuse emission of seven fields near the Galactic Center \citep{2021ApJ...913..115A} with an average position of  (l, b) = ($0^{\circ} .37257$, $-0^{\circ} .04588$).

\section{VTSCat at HEASARC}

Most of the information contained in VTSCat is also archived and available at the  HEASARC, which is familiar to a large part of the community of high-energy astrophysicists. 
Users can take advantage of the HEASARC through multiband data searches and VERITAS data can  be included in these searches. 
The main catalog at the HEASARC will be a source catalog including all sources detected with VERITAS. 
The SEDs and light curve data for individual sources will be in individual files linked to the main source catalog. 
As in VTSCat, the objects and sources are identified by their source identifiers (IDs).
All data fields along with their units are described in detail at \url{https://heasarc.gsfc.nasa.gov/W3Browse/all/verimaster.html}. 
Following VTSCat, the data files wrapped in FITS format contain lightcurve data (integrated gamma-ray fluxes vs time) and spectral flux points (including errors and upper limits). 
The data types and units for these data products are described in the FITS header.

\section{Outlook}

VTSCat will be updated regularly with new publications of the VERITAS collaboration.
Future structural updates of VTSCat may include multi-wavelength data presented in VERITAS publications including observations by e.g., radio, optical, or X-ray instruments, and may also allow to use tools like the SED builder provided through SSDC\footnote{\url{https://tools.ssdc.asi.it/SED/}}. 
The effort of disseminating all VERITAS results in digital format will  aid in multiwavelength analysis and input to the modeling of sources of high-energy gamma rays.
%



\acknowledgments

We thank Antara Basu-Zych and Alan Smale for their assistance in archiving the VERITAS catalog at HEASARC.

This research is supported by grants from the U.S. Department of Energy Office of Science, the U.S. National Science Foundation and the Smithsonian Institution, by NSERC in Canada, and by the Helmholtz Association in Germany. This research used resources provided by the Open Science Grid, which is supported by the National Science Foundation and the U.S. Department of Energy's Office of Science, and resources of the National Energy Research Scientific Computing Center (NERSC), a U.S. Department of Energy Office of Science User Facility operated under Contract No. DE-AC02-05CH11231. We acknowledge the excellent work of the technical support staff at the Fred Lawrence Whipple Observatory and at the collaborating institutions in the construction and operation of the instrument.
The material is based upon work supported by NASA under award number 80GSFC21M0002.

\facilities{VERITAS}

\end{document}